# Quantum Composites with the Functionality Defined by the Charge-Density-Wave Phase Transitions

**Zahra Barani**[1,*], **Tekwam Geremew**[1,*], Megan Stokey[2], Nicholas Sesing[3], Maedeh Taheri[1], Matthew J. Hilfiker[2], Fariborz Kargar[1,†], Mathias Schubert[2], Tina T. Salguero[3], and Alexander A. Balandin[1,†]

[1]Phonon Optimized Engineered Materials Center, Department of Electrical and Computer Engineering, University of California, Riverside, California 92521, USA

[2]Department of Electrical and Computer Engineering, University of Nebraska, Lincoln, Nebraska 68588, USA

[3]Department of Chemistry, University of Georgia, Athens, Georgia 30602, USA

We demonstrate a unique class of advanced materials – quantum composites based on polymers with fillers comprised of a van der Waals quantum material that reveals multiple charge-density-wave quantum condensate phases. Materials that exhibit quantum phenomena are typically crystalline, pure, and have few defects because disorder destroys the coherence of the electrons and phonons, leading to collapses of the quantum states. We succeeded in preserving the macroscopic charge-density-wave phases of filler particles after multiple composite processing steps. The prepared composites manifest strong charge-density-wave phenomena even above *room temperature*. The dielectric constant experiences more than *two orders of magnitude* enhancement while the material maintains its *electrically insulating* properties, opening a venue for advanced applications in energy storage and electronics. The results present a conceptually different approach for engineering the properties of materials, extending the application domain for van der Waals materials.

---

[*] Authors contributed equally

[†] Corresponding authors: fkargar@ece.ucr.edu (F.K.); balandin@ece.ucr.edu (A.A.B); https://balandingroup.ucr.edu/

Materials that reveal quantum phenomena are typically crystalline, compositionally pure, and have low defect densities.[1–9] It is known that disorder destroys the coherence of the electrons and phonons, leading to collapses of the quantum states.[10] Thus, one usually does not expect charge density wave (CDW) quantum condensate phenomena in disordered materials, *e.g.,* composites, that are mixtures of two or more materials with different properties. In this report, we describe remarkable *quantum composites* based on polymers with fillers comprised of a van der Waals quantum material, *1T*-TaS$_2$, that reveals multiple CDW condensate phases. Optical, thermal, and electrical characterization indicates that the macroscopic quantum states persist after composite processing steps. The macroscopic properties of quantum composites show signatures of the transition between the CDW quantum phases, which in this case exist even above *room temperature* (RT). The dielectric constant changes by orders of magnitude at the temperature of the transition between the two macroscopic quantum states in the filler materials.

Recent years have witnessed fast-growing interest in quantum and strongly correlated materials that show unusual electronic and magnetic properties, including metal-insulator transitions, superconductivity, and CDW phases.[8,9,11–17] The essential feature that defines strongly correlated materials is that the behavior of their electrons cannot be described effectively in terms of non-interacting entities. The models of the electronic structure of strongly-correlated materials must include electronic correlations to capture their properties. Examples of strongly correlated quantum materials include CDW materials of the transition metal dichalcogenide (TMD) family that have layered two-dimensional (2D) crystal structures. The 1*T* polymorph of the well-known TMD material, TaS$_2$, reveals the commensurate CDW (C-CDW) phase consisting of a $\sqrt{13} \times \sqrt{13}$ reconstruction within the basal plane at temperatures below ~200 K. Its Fermi surface is unstable, leading to the lattice reconstruction accompanied by a Mott - Hubbard transition that gaps the Fermi surface and increases the resistance.[14–16,18–22] The CDW phase transitions are not always described by the Peierls mechanism, do not necessarily involve the Mott - Hubbard strongly correlated effects, or rely on the Fermi surface nesting. However, despite the variety of CDW mechanisms, the strong electron-phonon coupling and correlations are considered essential features.[23-24] To observe the strongly-correlated or quantum phenomena, one normally must use samples with a high degree of crystallinity, high chemical purity, and low defect densities. Disorder destroys coherence and quantum correlations.[10] For this reason, considerable efforts have been directed at perfecting the growth of strongly correlated quantum materials by molecular beam epitaxy, chemical vapor deposition, chemical vapor transport (CVT), flux growth, and other methods.[25,26]

Multifunctional composites constitute a completely different type of materials obtained by mixing two or more different substances.[27–29] Polymer composites combine polymers, *e.g.*, thermosets or thermoplastics, with various fillers added to the base polymers to improve their performance for specific applications. The composites are obtained by mechanical mixing of the constituent materials followed by temperature or invasive chemical treatment, which often results in damage to the fillers. Desirable properties of composites include enhanced mechanical strength, increased thermal conductivity, dielectric constant, or electromagnetic shielding.[30–32] Functional composites usually require strong matrix–filler coupling, *e.g.*, polymer–graphene coupling is needed to increase the overall thermal conductivity of graphene composites.[33] To increase the matrix–filler coupling, the preparation of functional composites often involves surface functionalization, *e.g.* acid treatment, which results in more defects and disorder. Achieving the uniform dispersion of the fillers may require additional processing steps, adding to the filler damage.

We developed polymer composites with fillers comprised of exfoliated 1$T$-TaS$_2$, a van der Waals layered material with multiple CDW phases. We argue that such composites can be considered a distinct class of materials that have properties and functionalities defined by the quantum phenomena in the fillers. The uniqueness of 1$T$-TaS$_2$ is that it exhibits a transition between two CDW quantum phases at a temperature of ~350 K, notably above RT.[12–16,34–38] Synthesis of such composites is challenging due to mechanical defects to fillers during processing, filler agglomeration, filler oxidation, and possible intercalation of the matrix materials between the atomic planes of the van der Waals fillers. We succeeded in preserving the macroscopic CDW quantum states in the dilute composites and demonstrated the novel functionality of such composites defined by the transition between two quantum phases.

Importantly, the composites have a low filler loading – below and close to the percolation threshold – to preserve dielectric, *i.e.*, electrically insulating, properties of the material for its practical applications in energy storage, electronics, and optical coatings.

The 1$T$ polytype of TaS$_2$ is in the commensurate CDW (C-CDW) phase below ~200 K; it is in the nearly-commensurate CDW (NC-CDW) phase above this temperature; it undergoes a transition to the incommensurate CDW (IC-CDW) at ~350 K, and it becomes a normal metal phase above ~545 K.[12–16] Each phase transition is accompanied by lattice reconstruction, leading to modifications of the material's electrical properties. The C-CDW phase drives a Mott-Hubbard metal-to-insulator transition resulting in a Mott insulator state. Intercalation, doping, defects, optical excitation, applied electric field, and sample thickness can modify the transitions to CDW phases.[12–16,39] For example, when the thickness of a 1$T$-TaS$_2$ film falls below ~10 nm, the C-CDW–NC-CDW transition at ~200 K disappears, whereas the NC-CDW–IC-CDW transition at ~350 K persists.[40] Figure 1a,b shows a schematic of the crystal structure of 1$T$-TaS$_2$ and C-CDW domains in the NC-CDW phase of this material. The robustness of the NC-CDW–IC-CDW quantum phase transition, which happens above RT, motivated us to focus on this transition in examining the properties of quantum composites. The phase transitions above RT are also beneficial for practical applications.

The bulk 1$T$-TaS$_2$ crystals used in this work were prepared using the chemical vapor transport (CVT) method. The details of the synthesis procedures, optical microscopy images of as-grown crystals, and the results of X-ray diffraction measurements that confirm the phase and crystallinity of the synthesized materials, are provided in the Supporting Information, Figures S1-S3. The energy-dispersive X-ray spectroscopy (EDS) data from a representative crystal are presented in Figure 1c. The purple and yellow maps show the homogeneous distribution of Ta and S atoms, respectively; the image in the lower panel is the combined EDS map. Further analysis of the EDS data can be found in Table S1 and Figure S4. To verify the CDW properties of as-grown 1$T$-TaS$_2$, we exfoliated single-crystal multilayers, *i.e.*, flakes, placed them on Si/SiO$_2$ substrate, and fabricated Au contacts for the electrical measurements. The lateral dimensions of the flakes were in the range of 1 μm to 5 μm and the thicknesses were in the range from 10 nm to 40 nm. These dimensions approximately correspond to a typical 1*T*-TaS$_2$ filler obtained by the chemical exfoliation for the preparation of the composites. The details of fabricating the Ohmic contacts to 1*T*-TaS$_2$ samples for the electrical measurements have been reported by some of us elsewhere.[14,41,42] In Figure 1d we present the representative current-voltage (I-V) characteristics of a single-crystal few-layer of 1*T*-TaS$_2$ measured at room temperature. The hysteresis observed by applying the forward and reverse electric bias clearly reveals the NC-CDW to IC-CDW phase transition, in line with prior reports.[12,16,18–21,34–37,40–45] These results confirm that the filler material we used for the preparation of quantum composites is of high quality, with a well-defined and pronounced CDW phase transition.

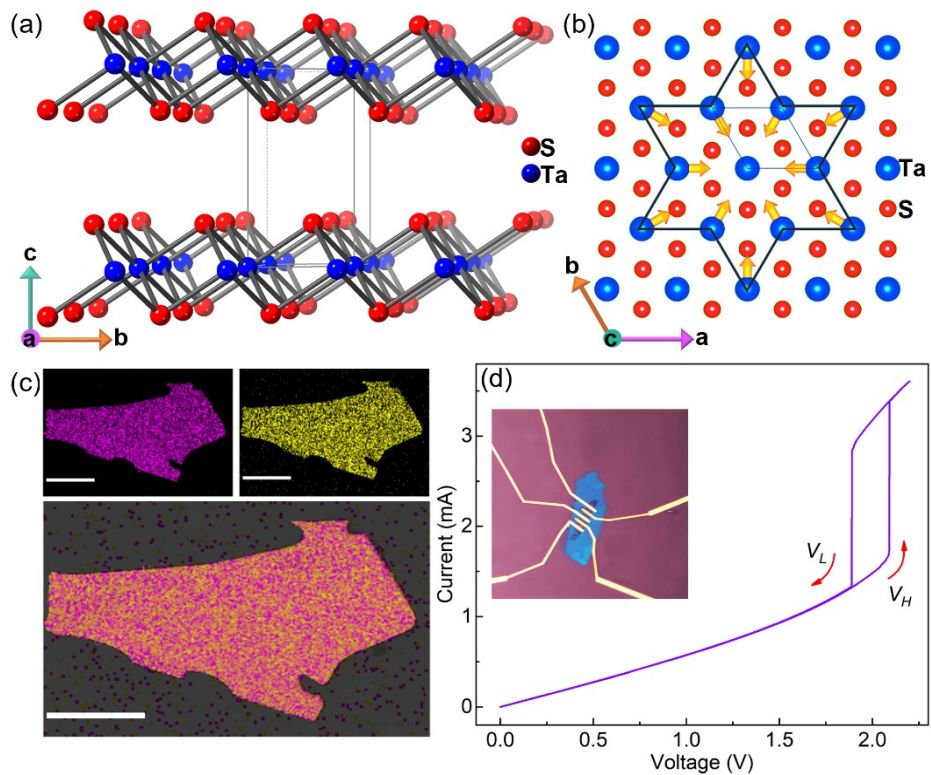

**Figure 1: Crystal structure and CDW quantum phase transition in as-grown 1T-TaS$_2$ materials** (a) Schematic of the crystal structure of 1*T*-TaS$_2$ showing its quasi-2D nature. (b) Crystal lattice reconstruction that accompanies the CDW quantum phase transitions, indicating the displacement of the Ta atoms. (c) EDS maps of an as-synthesized 1*T*-TaS$_2$ crystal showing the elemental distribution of Ta (purple) and Se (yellow) atoms. The combined EDS map is provided in the bottom panel. All scale bars are 500 μm. (d) Room-temperature current-voltage characteristics of an individual single-crystal flake of 1*T*-TaS$_2$. The hysteresis loop near the applied bias of 2.1 V corresponds to the NC-CDW to IC-CDW quantum phase transition. The arrows indicate the forward and reverse bias sweep. The inset presents an optical microscopy image of the individual 1*T*-TaS$_2$ flake of approximately ~5 μm length with the fabricated metal contacts for electrical measurements.

We have developed an original technique for the preparation of composites with 1$T$-TaS$_2$ fillers, focusing specifically on dilute composites with low loading, $\varphi$, of CDW fillers ($\varphi < 5$ vol. %). The filler loading of slightly below the percolation threshold ensures that the composites demonstrate unusual dielectric constant while remaining electrically insulating. Preserving the electric insulation properties of the composite is important for the intended applications. Figure 2a presents a scanning electron microscopy (SEM) image of the as-grown bulk material, confirming its quality. The CVT-synthesized crystals were exfoliated into filler particles with optimized dimensions through a liquid-phase exfoliation (LPE) procedure using *dimethylformamide* (DMF) solvent and bath sonication (Figure 2b). The exfoliated fillers were mixed with epoxy and *polyvinylidene fluoride* (PVDF) polymers to prepare flexible films (Figure 2c). More details of the composite preparation procedure are described in the Methods section and Supporting Information (Figure S5, Table S2).

The synthesis steps have been guided by physics and chemistry considerations and optimized to avoid excessive damage, oxidation, or other alteration to the fillers. A certain degree of control of the filler thickness and lateral dimensions can be achieved *via* a proper selection of the solvent and sonication time.[46,47] The fillers will always come with the distribution of sizes but one can still identify some average parameters.[48] We considered that in general the filler thickness above ~ 10 nm [14,40] and the lateral dimensions on the order of the CDW coherence length of a few μm are beneficial for preserving the CDW phase transitions. While the amplitude coherence length in CDW materials is small, on the order of ~10 nm, the phase coherence length can be up to ~10 μm. [1-3] The filler size and distribution across the thickness and along the surface of the prepared composites were analyzed using SEM inspection. Supplementary Materials Figure S6a-d shows cross-sectional and surface SEM images of a representative composite with *1T*-TaS$_2$ fillers. As seen in Figure S6c, the lateral dimension of the fillers varies from a few to ~40 μm, consistent with the filler distribution after chemical exfoliation.[48,49] We note that the exact distribution of the filler sizes is not crucially important in the context of this research. The filler loading fraction and proper dispersion in the matrix material are more important parameters. The SEM data confirms uniform dispersion of the fillers, without substantial agglomeration; the orientation of the fillers is random as intended.

The crucial step of creating quantum composites with CDW fillers is verification that the fillers retained their quantum condensate phases in the composites. We accomplished the verification *via* Raman and differential scanning calorimetry (DSC) measurements. Raman spectroscopy is a standard technique for monitoring CDW phase transition owing to its sensitivity to the crystal lattice distortions, reconstruction, phonon folding due to CDW periodicity, and the loss of translation symmetry in the IC-CDW phase.[50–54] The Raman spectra were accumulated in the conventional backscattering configuration at T ~ 93 K. The Raman laser spot size was ~1 µm. Experiments were conducted with the minimum laser power to avoid any laser-induced heating effects. Figure 2d shows Raman data for bulk 1$T$-TaS$_2$ crystal before and after LPE and for a dilute composite with 1$T$-TaS$_2$ fillers. In bulk 1$T$-TaS$_2$, (purple spectrum), the intense peaks in the range of 70 cm$^{-1}$ to 130 cm$^{-1}$,[41,42] as well as the low-intensity peaks in the range of 225 cm$^{-1}$ to 400 cm$^{-1}$ range,[51,52] are attributed to the zone-folded acoustic and optical phonons. These phonon branches appear because of Brillouin zone reconstruction in the C-CDW phase and disappear during the C-CDW–NC-CDW phase transition.[52] At 93 K, the bulk crystal and the fillers are all in the C-CDW phase. As can be seen, all these peaks are present in the spectra associated with the exfoliated material alone (blue curve) and when it is incorporated as a filler within the epoxy matrix (orange curve). The latter proves that the 1$T$-TaS$_2$ fillers preserve their CDW properties after chemical exfoliation and mixing with the polymer matrix.

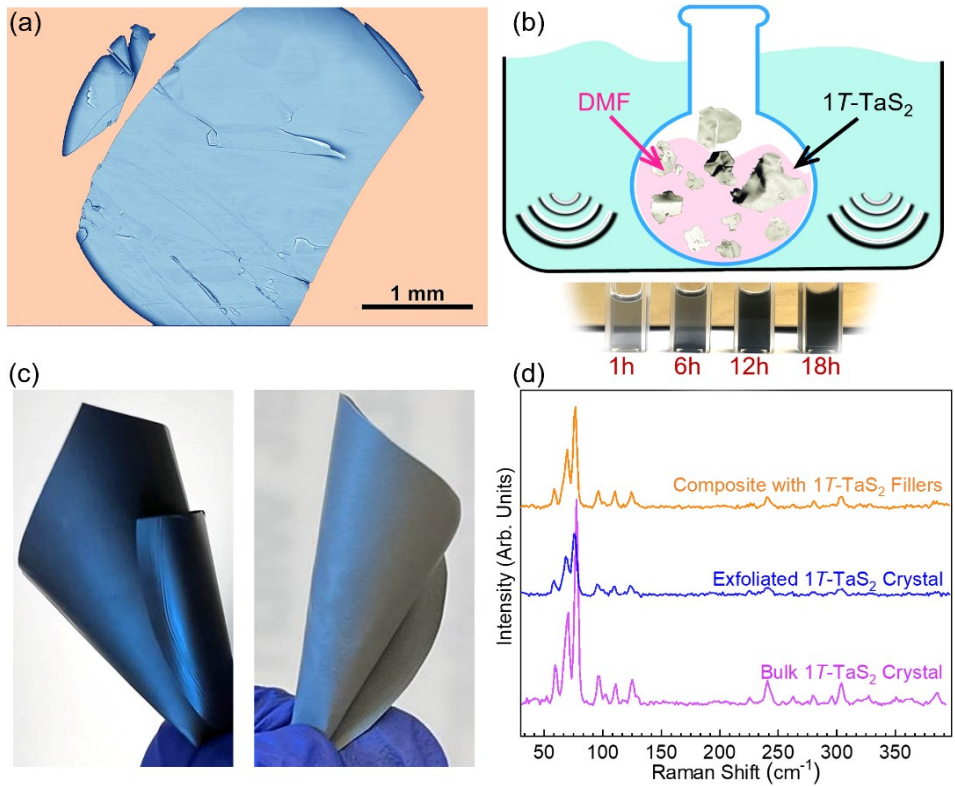

**Figure 2: Preparation of quantum composites with 1$T$-TaS$_2$ CDW fillers.** (a) SEM image (false color) of 1$T$-TaS$_2$ crystals synthesized by the CVT method. (b) Illustration of the chemical exfoliation method using low-power bath sonication in DMF solvent. The vials with DMF solutions of 1$T$-TaS$_2$ fillers are shown at different exfoliation times. (c) Flexible PVDF composite films with 1$T$-TaS$_2$ fillers (left) and dual 1$T$-TaS$_2$ and $h$-BN fillers (right). (d) Raman spectra of the bulk 1$T$-TaS$_2$, exfoliated fillers, and dilute epoxy composites with 1$T$-TaS$_2$ fillers under 488-nm laser excitation at T = 93 K. The Raman spectra reveal the presence of CDW phases after all processing steps.

To monitor the strongly-correlated effects in the composites directly, we conducted temperature-dependent Raman spectroscopy and differential scanning calorimetry (DSC). Figure 3a,b presents the evolution of the Raman peaks with temperature, $T$, for a pristine crystal and a dilute composite, respectively. The temperature-dependent Raman spectra for the liquid-phase exfoliated fillers are presented in the Supporting Information, Figure S7. The experiments were carried out during the heating cycle. At $T$~223 K, which corresponds to the NC-CDW–IC-CDW phase transition, the low wavenumber peaks, between 50 cm$^{-1}$ and 80 cm$^{-1}$, merge into one broad peak; and the other zone-folded acoustic and optical phonons disappear (Figure 3a). The magnitude of the peaks in the range of 100 cm$^{-1}$ to 130 cm$^{-1}$ decreases significantly. At ~363 K, which corresponds to the NC-CDW–IC-CDW transition, no pronounced Raman peaks can be resolved, confirming that all observed peaks were associated with phonon modes folding back from the reconstructed Brillouin zone edge in the C-CDW state. This is expected due to the loss of translation symmetry in the IC-CDW phase.[50] The evolution of phonon modes as a function of temperature is characteristic of the CDW phase transitions and can be used to determine whether the material preserves its CDW properties after the LPE and inclusion into a polymer matrix. Figure 3b shows Raman peaks of the dilute epoxy composite with 1$T$-TaS$_2$ fillers. The peaks follow the same evolution with temperature, confirming that the fillers undergo the same quantum phase changes.

Figure 3c,d shows the results of the DSC measurements of pristine crystal 1$T$-TaS$_2$, exfoliated 1$T$-TaS$_2$, and dilute epoxy composite with 1$T$-TaS$_2$ fillers, respectively (see Figure S8 for DSC data of PVDF composites). The results for the single crystal 1$T$-TaS$_2$ show three distinct peaks with the respective CDW transition temperatures of ~218 K, 275 K, and 353 K. These peaks are associated with the C-CDW–NC-CDW, 1T', and NC-CDW–IC-CDW phase transitions, respectively.[55] The peak attributed to the C-CDW–NC-CDW transition is detectable for the exfoliated 1$T$-TaS$_2$ but vanishes for the composite samples. The 1T' peak disappears in both the exfoliated fillers and composites. It was previously reported that the C-CDW–NC-CDW and 1T' transitions depend strongly on the film thickness, interlayer van der Waals interactions, and applied pressure.[10,36] Such transitions are known to be more fragile and vanish in crystalline 1$T$-TaS$_2$ films thinner than ~10 nm.[36] We also observe that the two peaks associated with these transitions fade away as we thin down the fillers during LPE or polymer mixing processing steps. The NC-CDW–IC-CDW transition is still clearly visible in the DSC data of the composites. The latter suggests that this transition remains strong in the CDW fillers after LPE and incorporation into the polymer host.

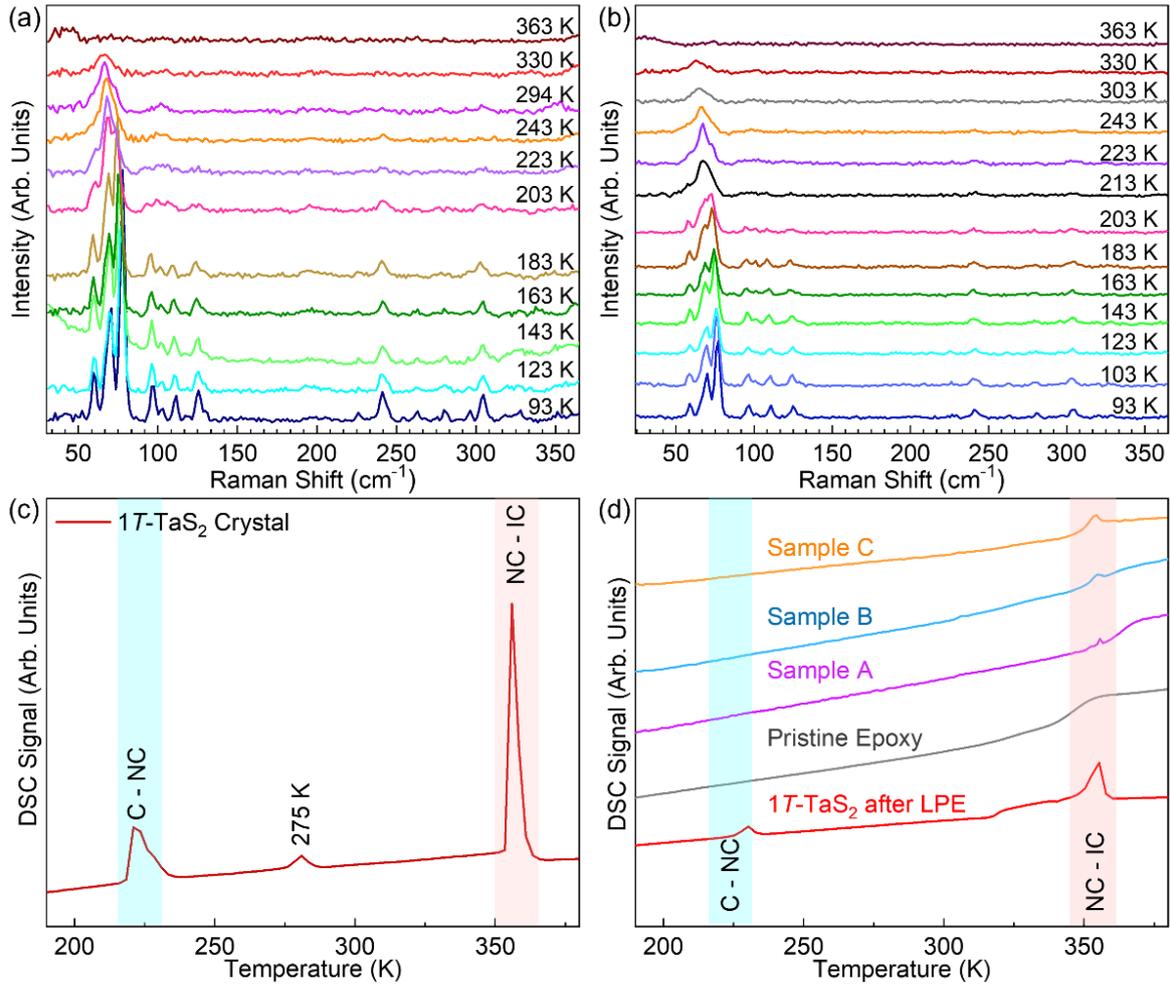

**Figure 3: Raman and DSC analysis of quantum composites.** (a) Raman spectra of single-crystal 1$T$-TaS$_2$ at different temperatures, varying from 93 K up to 363 K. (b) Raman spectra of dilute epoxy composite with 1$T$-TaS$_2$ fillers at different temperatures. Note that the spectra evolve similarly. The peaks in the ~100 cm$^{-1}$ to 125 cm$^{-1}$ range disappear at ~213 K as a result of the C-CDW to NC-CDW transition. All Raman peaks disappear after T~355 K as a result of the loss of translation symmetry in the IC-CDW phase. (c) DSC signal of the reference 1$T$-TaS$_2$ crystal. (d) DSC signals of the exfoliated 1$T$-TaS$_2$ in DMF (red curve), pristine epoxy (gray curve), and dilute composites with 1$T$-TaS$_2$ fillers. The filler loading for samples A, B, and C are 0.3 vol%, 1.3 vol%, and 1.4 vol%, respectively. Note that the DSC signature of the transition from NC-CDW to IC-CDW quantum phase is present in all samples.

A disordered material system that reveals CDW condensate phases at room temperature is already an important development and a conceptual change from the prior line of thinking. The quantum phenomena in composites become even more valuable if they affect the macroscopic properties of the composites and provide new functionalities. One of the unique characteristics of bulk crystalline CDW materials includes an anomalously high dielectric constant, $\varepsilon_1$.[56,57]

However, the utility of the unusual dielectric constant in brittle, bulk crystalline materials with metallic electrical conductivity is limited. Electrically insulating materials are more practically relevant for electrical energy storage or optical switching applications. We investigated the optical properties of pristine *1T*-TaS$_2$ and quantum composites using spectroscopic ellipsometry.[58] The details can be found in the Supporting Information (see Figure S9 for the ellipsometry data of bulk *1T*-TaS$_2$). Figure 4a,b shows the real, $\varepsilon_1$, and imaginary, $\varepsilon_2$, parts of the dielectric function of reference epoxy used in the composite preparation. The data do not show any strong changes. The temperature dependence of the dielectric constant of dilute composites is drastically different (see Figures 4c,d). Both $\varepsilon_1$ and $\varepsilon_2$ experience an abrupt jump at ~355 K, which is observed not only for the infrared wavelengths but also for the visible and UV light. This behavior is due to the properties of the CDW fillers, which undergo a transition from the mixed NC-CDW phase to a more metallic IC-CDW phase (see Supplementary Materials for $\varepsilon_1$ and $\varepsilon_2$ in pristine *1T*-TaS$_2$ crystal). It is important for practical applications that CDW quantum composites are diluted and maintain highly resistive, *i.e.*, electrical insulator, properties.

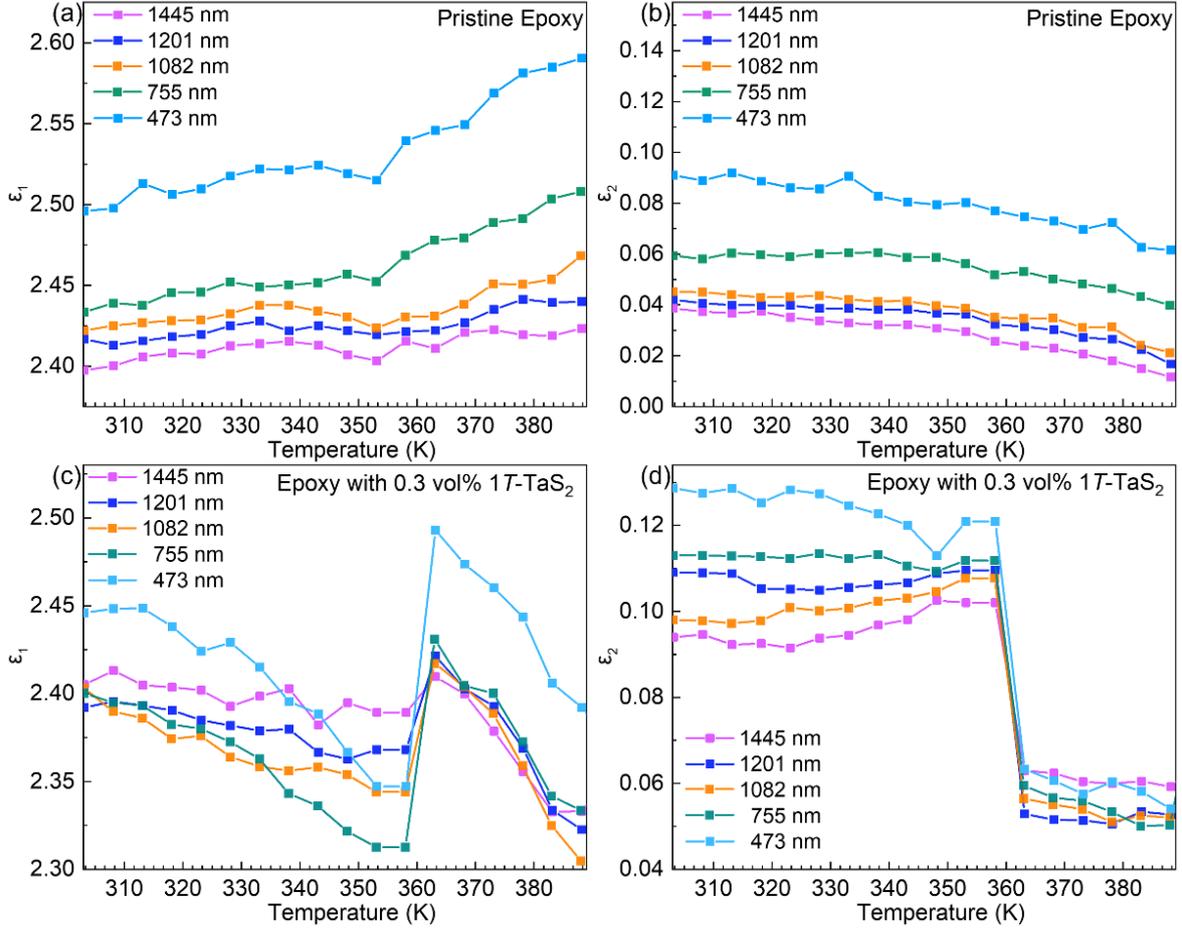

**Figure 4: Optical properties of quantum composites.** (a) Relative dielectric constant, $\varepsilon_1$, for pristine epoxy as a function of temperature at different wavelengths measured by spectroscopic ellipsometry. (b) The imaginary part of the dielectric constant, $\varepsilon_2$, as a function of temperature. (c) Relative dielectric constant, $\varepsilon_1$, for dilute composite with exfoliated $1T$-TaS$_2$ fillers as a function of temperature. (d) The imaginary part of the dielectric constant, $\varepsilon_2$, as a function of temperature for the composite. The data are presented for different wavelengths. Note an abrupt change in $\varepsilon_1$ and $\varepsilon_2$ at the NC-CDW to IC-CDW phase transition temperature of ~355 K in all cases.

To examine the DC characteristics of composites with PVDF base, we measured the capacitance of pristine PVDF and PVDF with $1T$-TaS$_2$ fillers as a function of temperature using the parallel-plate technique and LCR meter.[59] One should note that PVDF is a material of choice for capacitors. The results are presented in Figure 5a. The extracted $\varepsilon_1$ value for the pristine PVDF agrees with the literature.[60] Interestingly, the dilute composites with only 1.25 vol% and 4.5 vol% of $1T$-TaS$_2$ fillers reveal more than one and two orders of magnitude increase in $\varepsilon_1$ at the NC-CDW–IC-CDW transition temperature, respectively. In the control experiments, we measured the capacitance of the base polymer and a composite where $1T$-TaS$_2$

CDW fillers were replaced or mixed with the electrically insulating *h*-BN fillers. One can see in Figure 5b that $\varepsilon_1$ of the pristine PVDF and PVDF composites with *h*-BN fillers remain nearly constant in the examined temperature range. In the composite where *h*-BN and 1*T*-TaS$_2$ are mixed in the mass ratio of 1:2, the two orders of magnitude enhancement in $\varepsilon_1$ at the phase transition temperature is observed again. The experiments with the dual fillers, *h*-BN and 1*T*-TaS$_2$, demonstrate that the DC resistance of the composite can be fine-tuned and the leakage current further suppressed by the addition of electrically insulating secondary fillers.

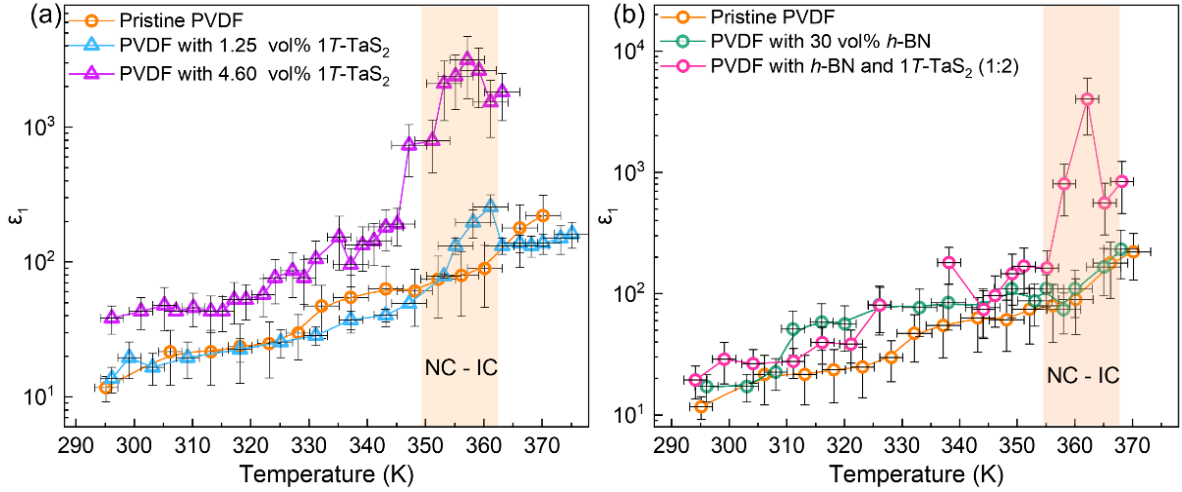

**Figure 5: Relative permittivity of quantum composites.** (a) Dielectric constant, $\varepsilon_1$, measured by the parallel capacitance method as a function of temperature for the pristine PVDF and PVDF composites with low-concentration of 1*T*-TaS$_2$ fillers. Note a strong enhancement of the dielectric constant at the CDW transition temperature. (b) The $\varepsilon_1$ data for two reference samples, *i.e.*, PVDF with electrically insulating quasi-2D *h*-BN fillers and PVDF with binary fillers of *h*-BN and 1*T*-TaS$_2$. The anomalous enhancement in $\varepsilon_1$ value is absent for the pristine PVDF and PVDF with only *h*-BN fillers.

The *colossal* dielectric constants, reaching $10^5$, were demonstrated in the single-crystal bulk CDW materials such as TaS$_3$, NbS$_3$, and (NbSe$_4$)$_3$I, at low temperatures. [56,57,61] These exceptionally high values were attributed to the extraordinary polarizability of CDW quantum condensate. In classical models, this was described with the large restoring force and oscillator strength of the pinned CDW distorted by the electric field of the electromagnetic wave. [56,57,61] All prior results were achieved for electrically conductive bulk crystals of metallic nature at temperatures below RT. We created an electrically resistive dilute composite with an exceptionally high dielectric constant, reaching $\sim 10^4$ at the temperature above RT. The effect was achieved due to the filler conversion to a more electrically conductive IC-CDW phase at the transition temperature. Even though CDW phases may be compromised in some filler particles due to oxidation, intercalation, or other degradation mechanisms, sufficient remaining particles preserve the CDW quantum condensate phase and provide the observed characteristics.

The main result of our research is the conceptually different approach for creating new functional composites that harness quantum phenomena in the fillers. We preserved the CDW quantum condensate properties of exfoliated 1$T$-TaS$_2$ filler material after its incorporation into polymer matrices and achieved two orders of magnitude enhancement in dielectric constant close to the phase transition, which occurs above RT. The enhancement in dielectric properties of composites with conventional fillers is usually explained by the increased polarizability of the medium under four different polarization mechanisms such as (i) electronic, (ii) ionic, (iii) interfacial, and (iv) dipolar polarization.[62] The dominant mechanism is defined by the type of filler and the host polymer. In the CDW materials, the intrinsic permittivity of the filler attains colossal values due to the exceptionally large polarizability of the CDW condensates close to the transition temperature.[56] Thus, the mechanism of increased polarization and consequently, $\varepsilon_1$, in the quantum composites is principally different from conventional composites.

We performed our proof-of-concept experiments with 1$T$-TaS$_2$ since this material has a robust NC-CDW – IN-CDW phase transition. One can envision that a similar approach can work with other CDW materials, which may produce even stronger effects. However, even now, we can provide a comparison with conventional composites in order to place our results within the general context of composite research. The relevant composite characteristics, which define the energy density of a dielectric capacitor are $\varepsilon_1$ and the dielectric breakdown field strength, $E_b$.[63] Many different types of fillers have been incorporated into polymer matrices to enhance $\varepsilon_1$ of the dielectric composites. These fillers include conventional ceramics, metals, carbon-based materials (e.g., graphite, nanotubes, reduced graphene oxide), semiconducting oxides, and ferroelectric materials, e.g., BaTiO$_3$.[64–69] BaTiO$_3$ has a high intrinsic dielectric constant, $\varepsilon_1 > 1000$, depending on the material's synthesis method and particle size and is widely used in nanodielectric composites.[69] It has been shown that the dielectric constant of PVDF-BaTiO$_3$ composites with filler concentration of $\varphi = 60$ vol% reaches to $\varepsilon_1 \sim 120$ at 1 kHz at RT.[68] Notably, this $\varepsilon_1$ value is achieved only at a much higher filler concentration compared to the loading used in this study for 1$T$-TaS$_2$. Polymer-based composites with metal and carbon-allotrope fillers at concentrations below the percolation threshold demonstrate high $\varepsilon_1$ because of increased interfacial polarization.[69] For example, epoxy-based composites containing silver nanoparticles with $\varphi = 22.3$ vol% show $\varepsilon_1 \sim 300$ measured at 1 kHz at RT.[70] Poly(vinylidene fluoride-*co*-chlorotrifluoroethylene) composites (P(VDF-CTFE)) with 0.51 vol% of reduced graphene oxide filler exhibits $\varepsilon_1 \sim 40$ at 1 kHz at RT. [71] Importantly, in the case of conductive composite with high loadings of fillers, the enhancement in $\varepsilon_1$ is overshadowed by the increased leakage current and poor voltage breakdown field strength.

The anomalously high dielectric permittivity of the polymer-based dilute composites with CDW fillers exceeds many previously reported composites and offers potential for a range of innovative applications related to variable capacitors, capacitive sensors, capacitors for energy storage, and reflective coatings. The temperature range at which the $\varepsilon_1$ value of the quantum composite drastically changes is close to the operating temperature of many electronic devices, increasing the feasibility of practical applications of such composites. Another aspect of the practicality of any new type of composite is production cost. Exfoliated 1$T$-TaS$_2$ and other CDW van der Waals fillers can be produced *via* the scalable and cost-effective LPE technique.[72] It is already being used for the processing of graphene fillers for thermal interface materials.[73] It is worth noting that 1$T$-TaS$_2$ preserves its unique CDW properties even in powder form. [74,75] Thus the use of 1$T$-TaS$_2$ powder for composite preparation may constitute another industrially-attractive method. A natural extension of the concept of quantum composites with CDW fillers would be the utilization of the one-dimensional (1D) van der Waals quantum materials that reveal CDW and topological phenomena.[76-80] Additional challenges that have to be dealt with in this research direction would include uniform dispersion of the high aspect ratio quasi-1D fillers and control of the polytypes of the synthesized materials; relevant materials, such as NbS$_3$, reveal polymorphism.[79,80] However, if implemented one can possibly develop composites with electrical and optical properties not achievable by other means.

In summary, we have demonstrated a new class of composites with properties defined by the quantum phenomena happening in the CDW fillers. These *flexible* and *mechanically robust* CDW quantum composites with the premise of *orders of magnitude enhancement* of the dielectric constant offer exciting prospects for innovative applications. We also developed a conceptually different approach for engineering the properties of such composites that opens new application domains for quantum and strongly-correlated materials.

**METHODS**

**Preparation of dilute composites with 1$T$-TaS$_2$ fillers**: The as-synthesized small 1$T$-TaS$_2$ crystals were added to dimethylformamide (DMF) solvent and exfoliated to few-layer 1$T$-TaS$_2$ fillers using low-power ultrasonic bath sonication (Branson 5510). Since 1$T$-TaS$_2$ is a van der Waals material with 2D motifs in its crystal structure it exfoliates into quasi-2D layers, *i.e.* stacks of a few atomic planes. The lateral dimensions of the fillers were in the range of a few micrometers after the LPE process. The thickness was in the range of a few nanometers to a few hundred nanometers. The exfoliation time can vary up to 18 hours. The resultant dispersion was centrifuged (Eppendorf Centrifuge 5810) at 12000 rpm for 20 minutes. To prepare composites, fillers were added to two different base polymers, epoxy, and PVDF. Note that epoxy is thermoset, whereas PVDF is thermoplastic. The samples were produced in the following steps. Firstly, pristine PVDF in powder was dissolved in N, N dimethylformamide (DMF ≥99%, Sigma–Aldrich) by magnetic stirring for seven hours at 360 K. Then, the high-concentration 1$T$-TaS$_2$ solution was added to the solution and stirred for another hour. After reaching the desired viscosity, the solution was drop cast into glass Petri dishes. Finally, by the complete evaporation of the DMF, the film was cured, gently removed, and put in the oven for two hours at 380 K. In epoxy resin composites, the exfoliated 1$T$-TaS$_2$ fillers after the last run of centrifuging were poured into glass Petri-dishes and were left on top of a hotplate for two hours, with the temperature set to 360 K. After three hours, the dried exfoliated 1$T$-TaS$_2$ fillers were collected, weighed and added to epoxy resin (bisphenol-A-(epichlorohydrin), molecular weight ≤700, Allied High-Tech Products, Inc.) and hardener (triethylenetetramine, Allied High-Tech Products, Inc.) with the mass ratio of 100 to 12, respectively. To have a homogenous composite, we used a high-speed shear mixer (Flacktek Inc.) at a speed of 300 rpm for 15 minutes. The mixture was vacuumed for 20 minutes to avoid any possible air bubbles. The final mixture was poured into round disk-shaped silicone molds and was left at RT for eight hours. In the end, the cured sample was left in the furnace for another two hours at 390 K. Finally, by calculating the exact density of each film and composite using an electronic scale (Mettler Toledo) the exact vol% of each sample was determined. Additional information on the preparation and characterization of dilute composites with quasi-2D fillers can be found in the Supplementary Materials.

**Differential scanning calorimetry measurements:** Differential scanning calorimetry (DSC Polyma 214, NETZSCH-Gerätebau GmbH, Germany) was used to determine the CDW phase transition temperatures of the samples. DSC can provide information regarding the changes in the physical properties of the material as a function of temperature and time. Small samples of pristine polymer, pristine crystal before and after exfoliation, and polymer with filler inclusions were placed in standard aluminum crucibles. The weight of all samples was selected to be in the range of 10 mg to 20 mg to provide a sufficient signal-to-noise ratio. Samples were examined in the dynamic temperature range of 77 K to 380 K with different heating rates of 5 K/min and 20 K/min. The DSC response of the samples was compared to that of the reference samples, which included an empty crucible and alumina. The CDW transitions appear as sharp peaks in the DSC spectra. The transition temperature and the heat flow associated with the transition can be derived by determining the onset and area of the respective peak. The variations in the heating rate of the experiments did not affect the results of our experiments, attesting to the high accuracy.

**Spectroscopic ellipsometry measurements:** Spectroscopic ellipsometry measurements were conducted using a rotating compensator ellipsometer (M-2000, J.A. Woollam Co., Inc.) in the spectral range 191 nm - 1687 nm. Samples were mounted on a heat cell sample stage (Heat Cell, J.A. Woollam Co., Inc.) and aligned at a 70° angle of incidence. The heat chuck was warmed in 5°C increments and left at each temperature for one minute from 30 °C - 120°C while data was acquired in situ every 30 seconds. Spectroscopic ellipsometry returns the complex reflectance ratio for each wavelength measured. From these values, the real and imaginary parts of the so-called pseudo-dielectric function can be directly calculated, assuming a two-phase model (substrate/ambient).[58] This pseudo-dielectric function closely resembles the dielectric function of the bulk material, provided that sample non-idealities such as overlayers or surface roughness are negligible.

**Acknowledgments**

A.A.B. was supported by the Vannevar Bush Faculty Fellowship (VBFF) from the Office of the Secretary of Defense (OSD), under the Office of Naval Research (ONR) contract N00014-21-1-2947 "One-Dimensional Quantum Materials" while developing the concept of quantum composites. T.T.S. supplied van der Waals bulk crystalline materials as part of the VBFF project collaboration. The work at UCR, which involved electrical measurements of $1T$-TaS$_2$, was supported, in part, by the U.S. Department of Energy Office of Basic Energy Sciences under contract No. DE-SC0021020 "Physical Mechanisms and Electric-Bias Control of Phase Transitions in Quasi-2D Charge-Density-Wave Quantum Materials." A.A.B. also acknowledges funding from the National Science Foundation (NSF) program Designing Materials to Revolutionize and Engineer our Future (DMREF) *via* a project DMR-1921958 entitled Collaborative Research: Data-Driven Discovery of Synthesis Pathways and Distinguishing Electronic Phenomena of 1D van der Waals Bonded Solids. The authors thank Sriharsha Sudhindra and Lokesh Ramesh at UCR for assistance with the testing of the materials and with the preparation of the Supplemental Information.

**Contributions**

A.A.B. conceived the idea of quantum composites, coordinated the project, contributed to the experimental data analysis, and led the manuscript preparation. Z.B. and T.G. prepared the composites, conducted Raman spectroscopy, specific heat and capacitance measurements, and contributed to data analysis. F.K. supervised the composite preparation, Raman spectroscopy, thermal, and capacitance measurements, and contributed to data analysis. N.S. synthesized and characterized bulk crystals by the CVT method. T.T.S. supervised the CVT material synthesis and contributed to the analysis of material characterization data. M.T. fabricated the devices, conducted current-voltage measurements, and contributed to composite characterization. M.S. and M.J.H. conducted spectroscopic ellipsometry. M.S. supervised the ellipsometry measurements and contributed to the optical data analysis. A.A.B. and Z.B. wrote the initial draft of the manuscript. All authors contributed to the editing of the manuscript.

**Declaration**

The authors declare no competing interests.

**Supplementary Information**

Supplementary information provides additional details of the composite synthesis and characterization data, including for bulk materials and reference samples.